# University of Minnesota
# CS 8701, Fall 2008

Paul Lesov

---

# Database Security: A Historical Perspective

**"How the database security controls adapted to threats over the last 30 years"**

# Table of Contents

# 1. Introduction

The importance of security in database research has greatly increased over the years as most of critical functionality of the business and military enterprises became digitized. Database is an integral part of any information system and they often hold sensitive data. The security of the data depends on physical security, OS security and DBMS security. Database security can be compromised by obtaining sensitive data, changing data or degrading availability of the database. Over the last 30 years the information technology environment have gone through many changes of evolution and the database research community have tried to stay a step ahead of the upcoming threats to the database security.

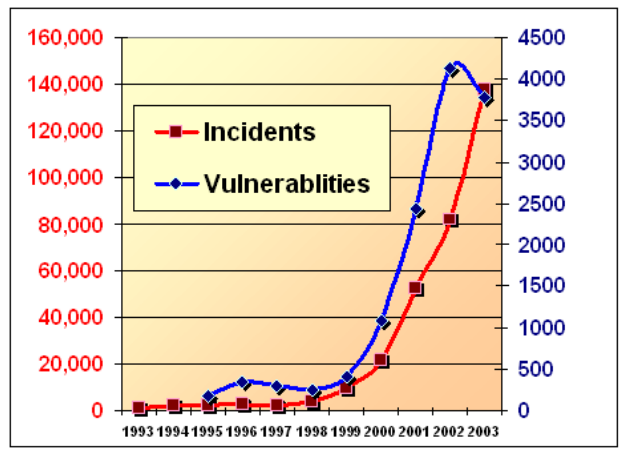

CERT 2003

On the first glance the CERT chart above suggests a losing battle. In reality database research is not too blame for the trend observed. The changing environment, non secure implementations and user errors is a main contributor to the explosive growth of the IT incidents and vulnerabilities.

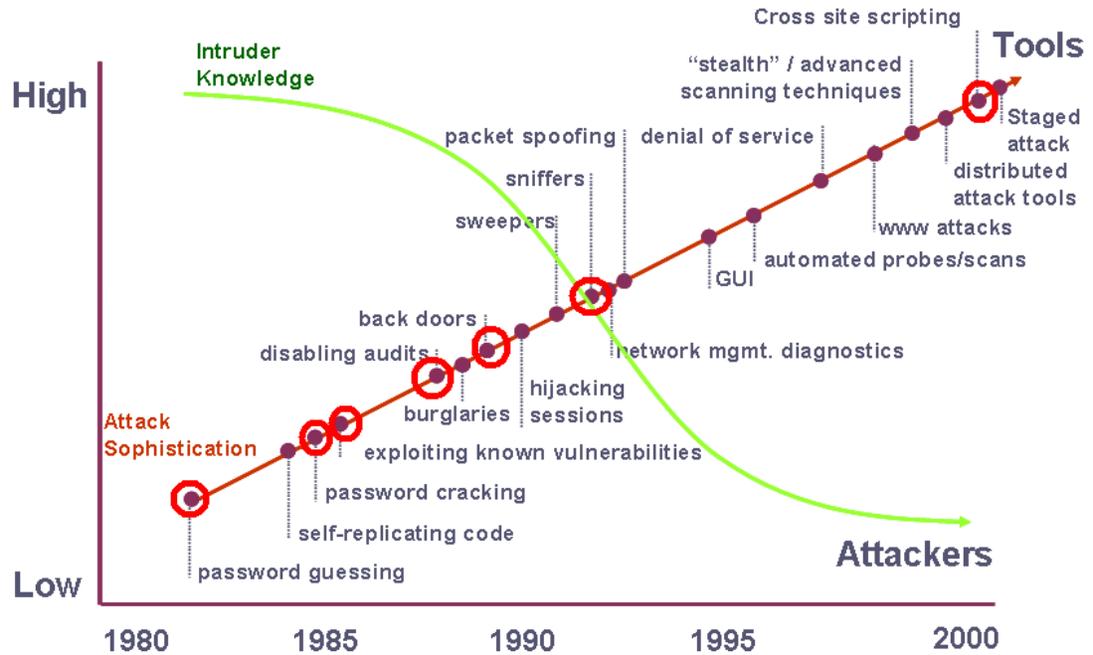

As systems become more modular and sophisticated attacker is presented with more vectors to conduct an attack. More over the attacker motivation have changed during the last 30 years. The technical barriers for executing an attack have also largely disappeared with easily available set of hacking toolkits as can be seen from this CERT 2000 visualization. Notice that we have marked with red circles the areas where database implementations were to blame for many of these security breaches. Often times vendors created back doors and default passwords on their password. For long time they did not support encryption for data in transit and the data could be eavesdropped over the network.

The database research community has thoughts about these issues long before they were address by the implementations. This paper will examine the different topics pertaining to database security and see the adaption of the research to the changing environment. Some short term database research trends will be ascertained at the conclusion.

# 2. The Early Years (pre 1980)

During this period database management system, especially the Relational model wasn't wide spread and its applications were handful. Government organizations like the Department of Defense (DOD) were the first to give high importance to security of data since their repository contained crucial data like the military data and census data. The commercial applications, very few though, were not greatly under threat since a formal security model was still under research. Organizations framed the security policies based on the few vulnerabilities identified. The physical and logical threats were identified.

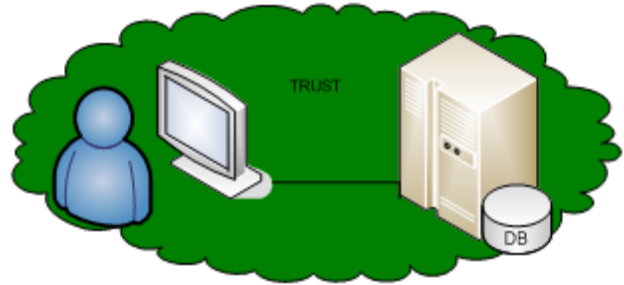

Physical threats were very well understood and security mechanisms were in place like securing the servers in authorized rooms. Logical threats were facing challenges and the security controls were not strong enough. Database systems were mostly static and offline. There was not much access of the systems across a wide network. Data wasn't stored across shared resources\repositories. The users accessed the systems over a trusted subsystem\environment relying mostly on the security of the underlying operating systems. Such trusted environment proved vulnerable to certain attacks like the trapdoors and Trojan Horse.

## 2.1.  Environmental Threats

It is worth mentioning the nature of the attackers during this period. Attackers were mostly sophisticated users who pursued attacks as a challenge.  Often, attackers used custom tools mostly developed by them or no tools at all. As a result, the attacks were not very destructive and sophisticated attack tools were not available to non sophisticated users.

Since the database systems were not applied to store data in shared resources, the major threats were in the inference of statistical data by users. Statistical databases were used to store data about the population or military information which were intended for aggregating useful information to be used by statisticians.

**Statistical Inference**:
The data stored in database system were queried to get the aggregate results. The query sets proved a threat since by simple observation of the query sets, information about individuals could be easily compromised. Based on the inference channel, the attacks can be classified as:

**Direct attacks**:

The attackers observed the queries and applied rules like *n-item k-percent rule* which says that an attribute of data which shows up for at least n items more than k percent of the result queried can be inferred.

**Indirect attacks**:
The aggregate results could be interpreted and *Trackers* could be applied on such results set to draw sensitive information. A Tracker is a simple formulation applied on the data like the intersection of query sets. A Tracker is analogous to a snooping tool, powerful though simple.

## 2.2. Mainstream Research

The main research was focused on the inference of sensitive data in statistical databases. The access control also gained momentum wherein models were proposed to restrict access to users based on the privilege of the users.

### 2.2.1. Inference

Active research was going on in understanding a Tracker and modeling controls to overcome such trackers. A common feature of a tracker is to control the set of records queried [1]. A control called Random Sample Queries [RSQ] was proposed in [1]. The basis of RSQ was to introduce uncertainty in the results so that the attacker cannot get reliable estimates of confidential data. The scheme works based on a probability measure for each record located. Whenever a record is located satisfying a characteristic C, the system determines if that record can be kept in the sampled query set. So, every time C is used, ideally same sample set should be used.

### 2.2.2. Access Control

Access control was one of the earliest security controls to be studied. While access control for file systems was well established an access control for DBMS did not fit well with that model. Access control for database was to be expressed in terms of logical data model with authorizations in terms of relations and tuples. It also had to be content aware to allow the system determine whether access should be granted based on the content of the data item. An important and benchmark work on access control was done for military applications by Bell and La Padula. This was imperative during this period because of the classified data stored in military applications. Data could belong to different levels and people accessing such data were given permissions on the level of the user and the level of data sought by the user. This came out to be called the famous lattice model or the Bell-LaPadula model(BLP). A lattice model consists of various levels and objects belonged to any of these levels based on the sensitivity of the data. Typical classification of military data was Top Secret, Secret, etc. The users were also mapped to one of the level in the user lattice. So, this model ensured that users could not read\write sensitive data higher or lower in the hierarchy. This model was developed to formalize the US Department of Defense multilevel security policy [4]. The BLP model was not completely applied in database management system until the early 1980s.

# 3. Growth to Maturity (1981-1990)

In the early 1980s, commercial applications using database systems started spreading. The attackers were motivated more out of the competition. Sophistication of tools used for attacks improved to enable attackers to exploit vulnerabilities with minimum knowledge of the target environment. Database systems were required to be stored in shared resources, increasing the need for security across the network and the users.

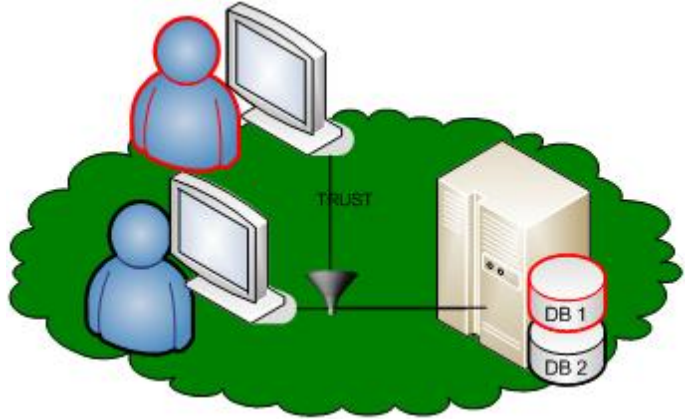

## 3.1. Environmental Threats

Most of the enterprises did not yet allow external connectivity but the number dumb terminal was replaced with client server model. The clients could not longer be fully trusted because these machines could be compromised. The database users were no longer necessarily technologist and were more likely to cause errors in the systems. This along with the fact that several departments were often sharing the same database required an access control.

## 3.2. Mainstream Research

The access control models which did not achieve any significant results continued to be the research focus while the statistical inference was also given importance which eventually diminished by mid 1980s. The following sections review the work done in these two areas.

### 3.2.1. Inference

Initial work during 1970s on inference proved futile since one of many arbitrary queries could break the controls. A breakthrough was achieved was the lattice model was applied which provided better understanding of inferences by the intruder. Suppression techniques and Perturbing techniques were applied to deny statistically useful information from getting leaked. Suppression techniques enforced restrictions on the set of allowable statistics. Different methods like suppressing the entire table or the individual cells were considered in this technique [2]. Perturbation techniques applied noise in the statistics either at the record level or the result level. The Random Sample Query (RSQ) discussed during the 1970s comes under this technique.

But not much of significant work on inference problems was followed in the mid 1980s. After early 1980s, work on statistical inference was stalled since statistical data was almost always subject to compromise through one of the several arbitrary ways [3]. Severe restrictions could lead to stale data not useful for statisticians

## 3.2.2. Access Control

The focus on security policies and mechanisms improved leading to increased momentum in area like the access control. Extensive research was carried on in access control and several models were proposed. Another area which also started seeing challenges was the multilevel database security. With models like the Bell-LaPadula model setting a benchmark in the security of classified multilevel data, Mandatory Access Control (MAC) models were researched. The Discretionary Access Controls (DAC) specified the rules which allowed the users, at their discretion, to create and delete objects, grant and revoke authorization for users.

DAC:

Access Matrix Model: The access rights (read, write, modify, delete, etc) were driven by this model which provides the permission for the subjects to act on an object (relations, views, attributes, etc). The rows represent subjects while columns represent objects and the entries represent access rights. Implementations based on this include authorization lists and capability lists. This model served as a powerful tool for implementing, studying and comparing access rights. [Denning et al.]. A relational implementation of this model was introduced in System R in which the access rights were maintained in a relation containing tuples of the form (Subject, Object, Access Right).

Query Modification: INGRES adopted this model which was similar to Access Matrix Model in addition to giving access control to views of each relation.
In System R, granting and revoking permissions was controlled using "copy flag" which had cascading revoking effect. SeaView project applied similar approach as System R for DAC but also applied not so found features then like authorization denials and precedence of rights.[2]

MAC

The shared databases required stringent policies for the users to get clearance in accessing data. Classified data could not be accessed by a user with a simple access right since the data could be highly sensitive to users based on the privilege of the users. As mentioned in [2], a mandatory policy formulates the following properties:
ReadP roperty: S can read d only if
secrecy-class(d) <= secrecy-class(S)
integrity-class(d) >= integrity-class(S)

WriteP roperty: S can write d only if
secrecy-class(S) <= secrecy-class(d) and
integrity-class(S)>= integrity-class(d).

No commercial system had implemented this model successfully in multi-level databases and research was still on. Research projects proposed few security models: I.P. Sharp model, the Hinke/Schaefer model, the TRW model, Navy surveillance model and SRI/Gemini Sea View Project [2]. Different system architectures which isolated security into kernel modules also played a vital role in providing the crucial mandatory controls. Some of them were decomposed database architecture and integrity lock architecture. Using decomposed database architecture required considerable changes on existing database before this architecture could be deployed to support multilevel relations and to interface with a security kernel. Integrity lock architecture did not require many changes on existing database but compromised on the assurance of mandatory security [2].

### 3.2.3. Encryption

Though access control model were developed and found to ensure security, there were always chances of those access controls to be bypassed leading to a breach. To enforce the second layer of security, data being stored in the repository could be modified and stored in an encrypted format. This idea gave way to research in the design possibilities of databases.

Two of such designs were Access Control Kernels and Encrypted Databases. Access Kernels were based on isolating and containing security policies inside separate modules. The downside of this design was that the value-dependent access restrictions were not possible. Trueblood et al [5] also provided modules based design in which security module was sandwiched between User & Application module and Storage\Retrieval module. These kernel architectures suffered from strong protection capabilities because of closeness to physical architecture than logical views\relations. The cryptographic technique of using keys to encrypt and store data was applied to achieve security. There were many restrictions and challenges like operations\computations on encrypted data, view-based protection, etc.

# 4. Rapid Data Expansion (1991-2000)

The digital environment during this era went through massive transformation driven by the rapid commercialization. The creation of Windows browser along with the realization that World Wide Web can be used for more than just information sharing, but actual commerce had a tremendous impact on the digital information exposure that made the security requirement of early 90's very different from the late 90's.

Another dominant development in computer science at this time was the programming methodology of Object Oriented Programming. These concepts have been added to many of the programming language of the time and it was natural for the database community to ascertain the efficient way to deal with complex object data. ODBMS (Object Oriented Database) took roots at this time as an alternative to RDBMS that is better suited for certain tasks commonly associated with OOP.

Towards the end of this era Online Analytical Processing (OLAP) became another important environmental component driving research. With large volumes of electronic data and digitalization of many aspects of the commerce efficient metric gathering became an important direction in research.

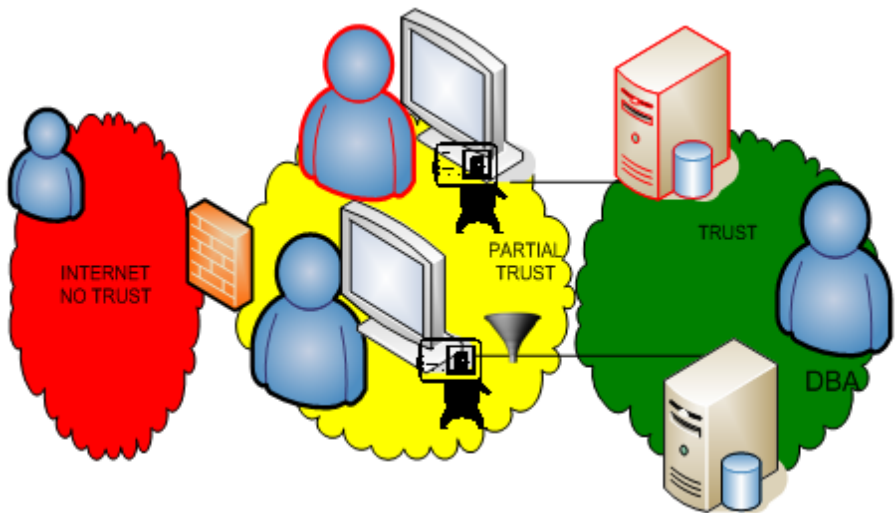

The widespread adaption of PC and spread of the Internet signaled the end of the isolated enterprises. Corporations wanted to have a presence to rip the benefits of WWW commerce and to advertise their products. Most put up perimeter defenses in the form of the firewalls to control which of their internal servers could be connected to from the un-trusted Internet and over which protocols. Enterprises widely adapted access controls for their internal database users. More and more users and application because users of the database servers. These users were often no longer SQL aware as application front ends because standard, shielding users from requiring technical expertise to use databases at this time.

While some departments used segregated servers many shared databases and required proper access control to maintain the principles of "need to know" and "least privilege". The Database administrators were still held under the assumption of full trust. Firewalls did not allow direct access to the database from any users on the untrusted Internet, which limited certain attack vectors.

## 4.1. Environmental Threats

Connectivity to the Internet brought about addition of untrusted connections. While firewalls provided protection against direct database attacks, the application front-end to the database were left vulnerable. Many applications did not provide adequate input validation and were susceptible to SQL Injections, which allowed attackers to insert harmful scripts into a database. This combined with the fact that many database vendors left database with default accounts and passwords often allowed attackers to compromises database from the Internet.

Earlier in the time period the attackers often were driven by publicity. Many countries did not yet enact strict computer crime laws and publicity was seen as an acceptable path to financial success as a top computer security consultant. The attackers therefore were often resulting to highly visible destructive tactics. It was increasingly critical for organizations to be able to reliability recover from these attacks.

Insider threat has grown as a concern as the number of individuals with access to the database grew with the propagation of online databases and computers at the workplace. The complexity and modularity of the database has also increased creating more attack vectors.

The introduction of data mining expanded the treat to individual privacy. Even with perfect access control and encryption the individual data can still be abused by a valid user. With database holding information from dozens of sources given an attacker an even better capability to inference. While data mining tools such as OLAP can be used for good such as counter-terrorist they can also be used to undermine the civil liberties of the individuals. Some government regulations such as HIPAA and FFIEC were established to safeguard individual privacy during this time.

## 4.2. Mainstream Research

Reviewing the database security related topics of this time reveals a heavy focus on refinement of access controls, privacy protection and efficient data encryption.

### 4.2.1. Privacy Protection

We have seen that Inference problem has a long history in database research. While up to 1990 these research focused on statistical databases the introduction of data warehousing change that. Just prior to our time period it was shown that the general inference problem was unsolvable [7][8]. In effect, the more inference controls were placed on the database it would became increasingly unusable. These early research did not have to concern itself with data mining from multiple sources. Many different techniques were introduced to deal with this problem that allowed for a compromise between system usability and privacy protection. Most fall in two groups. One approach is to modify data in some way so that a confidence of the sensitive association is reduced. Some techniques were research to estimate how much such perturbation impact the quality of the results. [9]. A data can also be sampled appropriately to be chosen in such a way as to many any inference from this data have a low degree of confidence. [10]

Another approach is an inference controller that is capable of detection and prevention of the inference attack. [11]

## 4.2.2. Access Control

While early work on integration database with mandatory and discretional access control produced break though such as System R Authorization Model, it was too rigid and suited for closed and controlled environments. The research of these models still continued. Mandatory access provided a higher level of security. Some research articulated a concept of polyinstantiation, or having multiple copies with different security levels of the same tuple in relation. [12][13]

Second half of this of this time period saw the emergence of the Extensible Markup Language (XML) as a new standard in describing data in documents moved over the Internet. It was realized early on that security was paramount in adaption of the new standard. XML access controls were developing along with parallel efforts on XML Encryption and Digital Signature. The nested structure of XML data required flexible protection object granularity. Most of the XML access control suggested therefore adapted to yes/no authorization and ex/implicit authorization that can be associated with a single document, such as a document specifying XML structure; document type definition (DTD).[14][15][16]

Also, the development of distributed object systems and e-commerce applications resulted in developments or object oriented access control. In fact, most of access control research at this time centered on role based and object oriented issues which we will now discuss in more detail.
.

### 4.2.2.1. RBAC

Role Based Access Control was often better suited for organizations than mandatory or discretionarily access control. RBAC is a type of discretionary access control where authorization administration is simplified around the notion of a role. Authorizations are given as need arises for an individual to perform a certain activity, rather than granted directly to user. Users are made up for roles thereby acquiring these authorizations.

RBAC is better suited for role hierarchies then other types of discretionary access controls as role-sub role relationships can be easily implied enabling authorization inheritance. Separation of Duties constraints can also be enforced easier since RBAC is centered on roles not users, thereby preventing a single user from receiving excessive authorizations.

Several RBAC models have been investigated [17]. RBAC model have been extended to add temporal dimension [18]. A standard have been developed for an XML based encoding of RBAC [16]. The development of the administrative models [18][19] and security analysis techniques[20] realized significant activity during this time. The implementation of these advanced features in the commercial products has been lagging. Many database vendors never did provide role based administrative functionalities into their products during this period and none provide temporal ones.

### 4.2.2.2.    ODBMS

Access control defined for relational databases do not suit well ODBMS. ODBMS access control system has to account for semantic modeling constructs often found in OO such as composite objects, versions and inheritance properties. With such complex data model a more fine grained protections system had to be developed for ODBMS. Both discretionary and mandatory access control systems were introduced to address this problem. The Orion object oriented DBMS was the first comprehensive discretionary access control model.[21].

Mandatory access control model for ODBMS are difficult to define due to the semantic richness of the OO. While research in this area was conducted [22]during this time little of practical significance was gained.

## 4.2.3.    Encryption

It was realized early that whichever access control the database contained the adversary still had a way accessing the underlying data by looking directly in the file system or underlying storage. However, for a long time cryptographic capabilities required a heavy cost burden for database operations and were not included in any implementations. With increases in processing speed supporting encryption of stored data became feasible in early 2000. At the same time several cases of adversaries and white hat hackers showed how easy it was to obtaining information from old discarded corporate hard drives. For databases supporting encryption of raw data created an additional issue for database optimization techniques, such as indexing. For example, if a column of a table containing sensitive information was encrypted the entire table would have to be scanned to provide a result to a query. Work on finding order preserving encryption for different data types is an active area of research.[23]

Data base federation allows integration of several autonomous database to act as a transparent single database. Queries can traverse from one database to other and often full trust can not be assumed between independent instances. A query may need to be performed without plaintext data being available to the database executing the query.[24]

One of the key aspects of security is the separation of duties and dual control. For a long time the databases were considering and adversary as an external entity being able to compromise data by subverting access control or via statistical inference. For a long time internal root user of DBA was considered to be fully trusted. Access control was not sufficient by itself to address to issue of DBA being able to exercise complete control over the data residing in the database. Early work concentrated on database implementations simply separating audit function from the DBA sphere of influence. Solutions often allowed logging all changes to the database in a stand along log file to which DBA had limited access. While helping with post incident detection and investigation this approach however did prevent the attack. Encryption provide a way to encrypt data in the database and store an encryption key external to the database thereby preventing the DBA from accessing the data [25].

# 5. Data Security Consolidation (post 2001)

This period thought data becoming ubiquitous. The new data types such as spatial and continuous sensor data entered the database field.

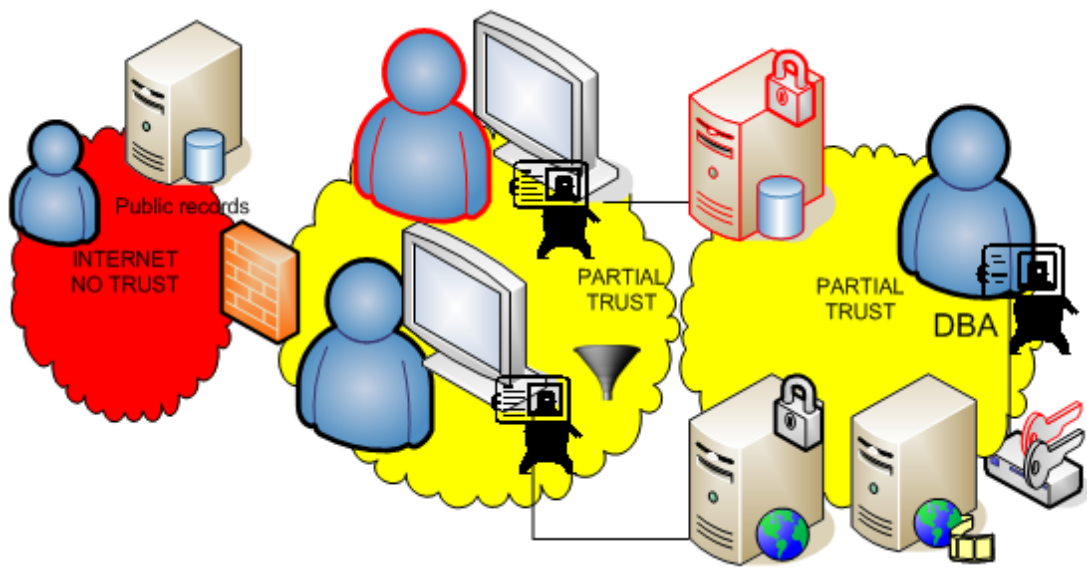

Personal information about individual became readily available through public records and via social Web 2.0 phenomena. The overall network has not changed much from the earlier decade fundamentally. Encryption became more common for the data not only in transit but at rest. User authentication methods have also been upgraded due to the threat posed by exploitation of the web applications up fronting critical data systems. Many institutions now use one time password devices and device fingerprinting technologies to more securely authenticate their users.

## 5.1. Environmental Threats

This period saw the emergence of information security attack as a large multi-billion businesses. Attackers have built sophisticated command and control networks running on millions of compromised computers. Attackers have also used sophisticated security techniques such as BitTorrent protocol and encryption to build resiliency and security around their perimeter. Attackers have also attempted to minimize any impact on the compromised system to limit the possibility of detection. Many tools are available today to conduct such attacks or utilize an attack service for a fee. The attackers have developed sophisticated business plans and many different avenues to profitability.

## 5.2. Mainstream Research

Research in access control have continued incorporating new data types such as spatial and sensor network data. With continues queries exhibited by sensor network access control could no longer make an authorization determination prior to query execution. It had to be able to account for temporal dimension as well to be effective.

Digital Right Management research also became prominent at the time with marking and pattern based detecting media data types in order to prevent intellectual data piracy.

### 5.2.1. Privacy Protection

Not all privacy concerns are driven by inference attacks. There was also research in anonymization of the data to protect privacy in the cases where the data is to be shared with the third party. [26]. Also, some suggested approaches to design database specifically tailored to support standard privacy policies such as the ones expressed in P3P. Paper [27] introduced a concept of Hippocratic databases that could support HIPPA requirement among others. These approaches use some privacy metadata consisting of privacy policies and privacy authorization tables. Paper [27] proposes a special attribute "purpose" to each table to perform privacy checking during query processing. Each query is issued with the purpose and the system checks the user, content and purpose based on privacy authorization table. Efficient are retention are some of the problem surrounding such approach and remain unresolved.

### 5.2.2. Access Control

More recently database systems have seen the emergence of new data types that require their own security controls. Restricting availability of string data type to a user is very different from restricting access to a specific view on a map grid if we take spatial data for example. There is an active research area of access control for spatial data with RBAC adoption to spatial data [28]. Some work is done on discovering mathematical authorization modes that would fit well with geographic data [29].

### 5.2.3. Encryption

We have yet to find any research of efficient encryption on this data in storage to counteract the threat posed by bypassing access controls and reading this data directly from storage media. It would appear that methods need to be found that could be used to secure this data but still allow efficient R-Tree indexing. Performing geometry encryption in such a way that operands such as intersect can be conducted on encrypted data may be the subject of future research.

### 5.2.4. Attack Detection

With attackers attempting to minimize impact on the target system researches had to tackle a problem of detecting a compromise. A side channel attack detection research is ongoing and many intrusion detection products have included database modules.

# 6. Conclusion and Future Trends

It is important to also note that many problems with securing data stored in the database is not due to the lack of research but lacking security in implementation of the database product or an application front ending the database. The shift from full trust to partial trust was driven in part by natural tendency to not provide full trust to anyone single individual based on dual control principle but also due to the inability of the users to keep their own PC computers secure and database frontend not being able to detect malicious attacks such as SQL injections.

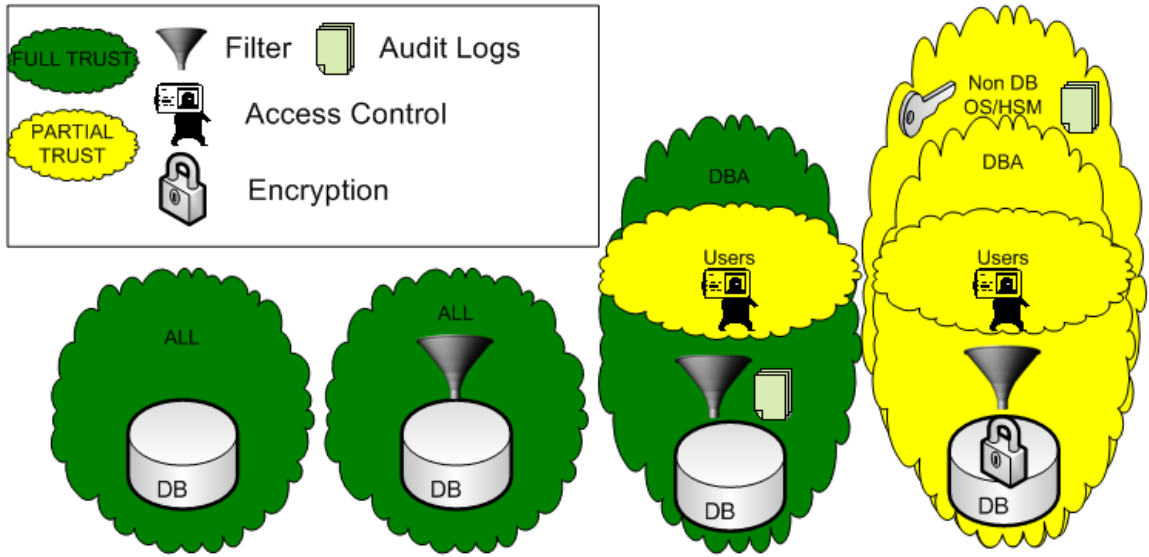

Time 1970→

   While intense focus on database security in the last 30 years created some welcome improvements it was still outpaced by the threat growth.  Data security concerns are evolving with new requirements emerging year to year.  The amount to data and the complexity and modularity of the databases is grown fast over time.  The recent decentralization process of outsourcing data processing creates a difficult problem for database security.  Intellectual Property Rights create a need for specific techniques for storing and marking this data and appropriate providing access to it.  New data types used in spatial and sensor network database require new ways to provide access control and encryption.  The semantic web outlook suggests the outsourcing trend to continue with related requirement for maintaining confidentiality and integrity of these query communications.

# 7. References

[1] Denning, D. E., Denning, P. J., Schwartz, M.D. 1979. *The tracker: a threat to statistical database security*. ACM Trans. Database Syst. 4(1): 76-96

[2] Denning D.E., *Database-Security Annual Review Of Computer Science*. 1988; 3: 1-22

[3] http://en.wikipedia.org/wiki/Statistical_database

[4] http://en.wikipedia.org/wiki/Bell-LaPadula_model

[5] Trueblood, R. P., Hartson, H. R., Martin, J. J. 1983. *MULTISAFE: A Modular Multiprocessing Approach to Secure Database Management*. ACM Trans. Database Syst. 8(3): 382-409

[6] C. T. Yu , F. Y. Chin, *A study on the protection of statistical data bases*, Proceedings of the 1977 ACM SIGMOD international conference on Management of data, August 03-05, 1977, Toronto, Ontario, Canada

[7] W.Jonge "Compromising statistical databases responding to queries about means", ACM Transactions on Database Systems, Volume 8, Issue 1, Marc h1983

[8] B.Thuraisingham "Recursion Theoretic Properties of the Inference Problem," MITRE report MTP291, June 1990

[9] S.Oliveira, O. Zaiane "Privacy Preserving Frequent Itemset Mining", Proc. IEEE ICDM Workshop, 2002

[10] C.Clifton "Using Sample Size to Limit Exposure to Data Mining", Computer Security, vol 8, no.2, Nov 2000

[11] B.Thuraisingham "Data Warehousing, Data Mining and Security" IFIP Database Security Conference, July 1996

[12] S.Jajodia, R.Sandhu, B.Blaustein "Solutions to the Polyinstantiation Problem", Information Security: An Integrated Collection of Essays" vol 1, IEEE CS Press, 1994

[13] R.Sandhu, F.Chen "The Multilevel Relational Data Model", ACM Trans. Information and System Security, vol 1, 1998

[14] E.Bertino, E.Ferrari "Secure and Selective Dissemination of XML Documents", ACM Trans", Information and System Security, vol 5, 2002

[15] A.Gabillon and E.Bruno "Regulating Access to XML Documents", Working Conference Database Security, 2001

[16] H.He, R.K.Wong "A Role-Based Access Control for XML Repositories", WISE 2000

[17] D.Ferraiolo, R.Chandramouli, R.Kuhn "Role Based Access Control", Artech House, 2003

[18] E.Bertino, P.Bonatti, E.Ferrari "TRBAC: A Temporal Role Based Access Control", Information and System Security, 2001

[19] J.Crampton, G.Loizou "Administrative Scope: A Foundation for Role-Based Administration", Information and System Security, 2003

[20] N.Li, M.Tripunita "Security Analysis in Role-Based Access Control", ACM SACMAT 2004

[21] F. Rabitti, W.Kim, D.Woelk "A Model of Authorization for Next Generation Database System", Database Systems, 1991

[22] M. B. Thuraisingham, Mandatory Security In Object-Oriented Database Systems, Conference Proceedings On Object-Oriented Programming Systems, Languages And Applications, P.203-210, October 02-06, 1989, New Orleans, Louisiana, United States

[23] R Agrawal, J Kiernan, R Srikant, Y Xu "Order preserving encryption for numeric data**" Proceedings of the 2004 ACM SIGMOD international conference

[24] P.Devanbu, M. Gertz, Chip Martel "Authentic Third-party Data Publication", Conference on Database Security , 2000

[25] Ueli Maurer, "The Role Of Cryptography In Database Security", Proceedings Of The 2004 Acm Sigmod International Conference On Management Of Data, June 13-18, 2004, Paris, France

[26] L. Sweeney "Achieving k-Anonymity Privacy Protection Using Generalization and Suppression", Uncertainty, Fuzziness and Knowledge-Based Systems, vol 10, 2002

[27] R. Agrawal, J. Kiernan, R. Srikant, And Y. Xu. "Hippocratic Databases" In 28th Int'l Conference On Very Large Databases, Hong Kong, China, August 2002

[28] F. Hansen, V.Oleshchuk "Spatial role-based access control model for wireless networks", Vehicular Technology Conference, 2003

[29] R. Thomas, R. Sundhu "An Authorization Model for Geospatial Data", IEEE Transactions and Dependable and Secure Computing, vol 1 2004